\documentclass[aip,jcp,reprint]{revtex4-1} 
\usepackage{graphicx}
\usepackage{enumitem}
\usepackage{mathtools}
\usepackage{graphicx}   
\usepackage{color}
\usepackage{ulem}
\usepackage{caption}
\usepackage{subcaption}
\usepackage{comment}
\usepackage{bm}
\usepackage{array}
\usepackage{booktabs}
\usepackage{multirow}

\usepackage{amsmath}

\usepackage{booktabs}

\begin{document}
\title{Accelerating all-atom simulations and gaining mechanistic understanding of biophysical systems through State Predictive Information Bottleneck}

	\author{Shams Mehdi}
 \affiliation{Biophysics Program and Institute for Physical Science and Technology,
 University of Maryland, College Park 20742, USA}

\author{Dedi Wang}
 \affiliation{Biophysics Program and Institute for Physical Science and Technology,
 University of Maryland, College Park 20742, USA}
 
\author{Shashank Pant}

\affiliation{Center for Biophysics and Quantitative Biology, Beckman Institute for Advanced Science and Technology, University of Illinois at Urbana-Champaign, Urbana, IL, 61801, USA.
\\
Current address for S.P.; Loxo Oncology {@} Lilly, Boulder, CO}

 \author{Pratyush Tiwary\footnote{Corresponding author.}}
 
 \email{ptiwary@umd.edu}
 \affiliation{Department of Chemistry and Biochemistry and Institute for Physical Science and Technology,
 University of Maryland, College Park 20742, USA.}

	\date{\today}
	
	\begin{abstract}

An effective implementation of enhanced sampling algorithms for molecular dynamics simulations requires \textit{a priori} knowledge of the approximate reaction coordinate describing the relevant mechanisms in the system. Here we demonstrate how the artificial intelligence based recent State Predictive Information Bottleneck (SPIB) approach can learn such a reaction coordinate as a deep neural network even from under-sampled trajectories. We demonstrate its usefulness by achieving more than 40 magnitudes of acceleration in simulating two test-piece biophysical systems through well-tempered metadynamics performed by biasing along the SPIB learned reaction coordinate. These include left- to right- handed chirality transitions in a synthetic protein (Aib)$_{9}$, and permeation of a small, asymmetric molecule benzoic acid  through a synthetic, symmetric phospholipid bilayer. In addition to significantly accelerating the dynamics and achieving back-and-forth movement between different metastable states, the SPIB based reaction coordinate gives mechanistic insight into the processes driving these two important problems. 
\end{abstract}

	\maketitle
	
\section{Introduction}
\label{sec:introduction}
Over the years, a variety of experimental methods have been developed allowing in-depth investigations of biomolecular systems.\cite{serdyuk2017methods}  However, a successful application of these tools to gain new insight is often not possible due to system complexity and limitations of the adopted tool itself.\cite{renaud2016biophysics} On the other hand, advances in computational hardware and efficient algorithm design have enabled implementing physical laws in a simulated world that allows the examination of these systems from a different perspective. Molecular dynamics (MD) is one such well established computational technique that generates a trajectory describing the time evolution of a system in all-atom, femtosecond resolution.\cite{karplus2002molecular} These trajectories can be analyzed to gain new thermodynamic and mechanistic insights. These insights can arguably be best encapsulated through the reaction coordinate (RC), which is the most informative mechanistic degree of freedom describing a system. The RC can differentiate between relevant metastable states and the pathways for moving between them. However, interpreting the huge amount of data produced by MD to identify the RC is not straightforward, and many methods have been proposed for this purpose.\cite{best2005reaction,ma2005automatic,peters2006obtaining,nadler2006diffusion,rohrdanz2011determination,mardt2018vampnets}

Additionally, complex and practically relevant problems can involve rare events and take place in time-scales that are still unreachable in MD. To overcome this limitation, a variety of enhanced sampling algorithms such as umbrella sampling, replica exchange MD, metadynamics, weighted ensemble and several others have been developed.\cite{torrie1977nonphysical,barducci2008well,sugita1999replica,zuckerman2017weighted,votapka2017seekr} A successful application of most of these methods can depend on the knowledge of the system's RC.\cite{bussi2015free} However, for practical problems it can be difficult to know the RC \textit{a priori} and a deduction of RC from MD simulations requires good sampling. This challenge clearly illustrates the need for developing novel algorithms to interpret MD results to learn the RC and tackle the rare event problem. 
State Predictive Information Bottleneck (SPIB) \cite{wang2021state} is one such recent method that enables learning the RC in systems with an arbitrary and \textit{a priori} unknown number of metastable states. It belongs to the Reweighted Autoencoded Variational Bayes (RAVE) \cite{ribeiro2018reweighted} family of methods. SPIB employs the concept of information bottleneck from information theory to approximate the RC of a system and is built on the principle of using minimal information from the past to reliably predict the state of the system at a future time. As shown in Ref. \onlinecite{wang2021state} such a state predictive information bottleneck approximates the perfect RC as given by the committor.\cite{bolhuis2002transition}

\begin{figure*}[t!]
    \centering
    \includegraphics[width=\linewidth]{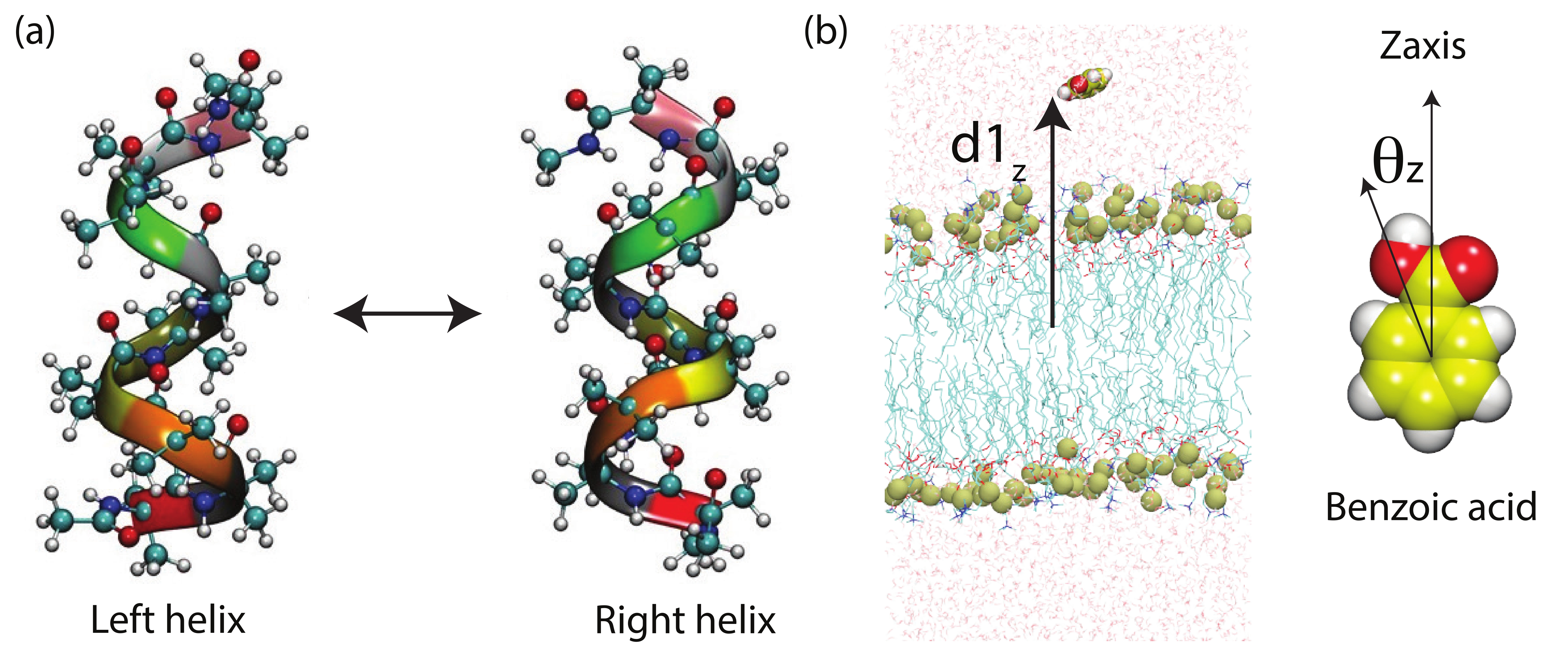}
    \caption{\textbf{Schematic representation of the systems studied in this work}: (a) Chiral transitions in (Aib)$_{9}$, (b) permeation of benzoic acid (BA) through a symmetric phospholipid bilayer. For (Aib)$_{9}$ we considered different dihedrals as order parameters (OPs). For the BA permeation through lipid bilayer, our OPs were built from two key physical quantities highlighted here: (i) membrane-BA Z distance, $d1_Z$ and (ii) angle between BA and Z axis ($\theta_Z$). Further details of OPs are given in Sec. \ref{sec:System_OP_details}.}
    \label{fig:system}
\end{figure*}
 
In this work, we demonstrate how the RC as learnt through SPIB performed on short, under-sampled trajectories can be used as a biasing coordinate in enhanced sampling, allowing significant and nearly automated acceleration of protein conformational dynamics and small molecule permeation through biological membranes. \cite{cardenas2012unassisted,shinoda2016permeability} Here we use SPIB with well-tempered metadynamics, but the protocol should be more generally applicable to other enhanced sampling methods.\cite{barducci2008well,tiwary2016review,sugita1999replica,zuckerman2017weighted,faradjian2004computing,defever2019contour,valsson2016enhancing,tiwary2013metadynamics,votapka2017seekr} Additionally, we show how SPIB can be performed on the biased trajectories to interpret them.

We demonstrate this SPIB-metadynamics protocol on two test-piece systems. First, we apply SPIB on $\alpha$-aminoisobutyric acid$_{9}$ (Aib)$_{9}$ which is a 9-residue synthetic peptide with alpha helical secondary structure (Fig. \ref{fig:system}(a)). On long time-scales the system is achiral, but on shorter time-scales the system undergoes left to right-handed chiral transitions and vice-versa.\cite{sittel2017principal,biswas2018metadynamics} Due to the time-scale limitations of unbiased MD, calculating the relative stability of fully left-handed and fully right-handed conformations of this system through this approach represents a computationally difficult task. Secondly, we study the permeation of a small asymmetric compound through a synthetic, symmetric phospholipid bilayer constructed from pure DMPC (1,2-dimyristoyl-sn-glycero-3-phosphocholine) lipids.\cite{lee2016simulation} The protonated benzoic acid (BA) molecule has a polar region that interacts with the hydrophilic head groups of DMPCs (Fig.~\ref{fig:system}(b)). For both systems we show how SPIB clearly helps in accelerating and making sense of the molecular dynamics. Additionally, we demonstrate different strategies involving the utilization of multiple independent trajectories as well as the initialization of the protocol (Sec. \ref{sec:Results}) for implementing SPIB. We thus believe this work represents a step forward in practical and automated use of AI-augmented enhanced sampling simulations for studying complex biomolecular problems.


\section{Methods}
\label{sec:Methods}
\subsection{State predictive information bottleneck (SPIB)}
\label{sec:spib_overview}
SPIB uses an information-bottleneck-based protocol\cite{alemi2016deep} to learn the RC, which then can be used to iteratively guide an appropriate enhanced sampling scheme. Consider a bio-physical system characterized by a set of order parameters (OPs) $\bm{X}$ and some metastable states $\bm{y}$. $\bm{X}$ can be amino acid dihedral angles for (Aib)$_{9}$ and specific distances, and angular coordinates for BA-DMPC as discussed in Sec. \ref{sec:System_OP_details}. The state label, $\bm{y}$, could be composed of the fully left-handed state and the fully right-handed state for (Aib)$_{9}$, and the bound state and unbound state for BA-DMPC. We describe these in detail in Sec. \ref{sec:Results}. As the number and location of such states are usually unavailable \textit{a priori}, SPIB only requires an initial assignment of state labels to launch its training process, and then it will automatically refine the assignments in an iterative manner. In SPIB, the RC is defined as the predictive information bottleneck $\bm{z}$ which carries the maximal predictive power for the future state of the system. In practice, a non-linear ANN encoder first converts the high dimensional input data into a low dimensional RC representation. Then, the ANN decoder classifies this RC space into different metastable states. In contrast to a RAVE \cite{ribeiro2018reweighted} decoder predicting the entire input space, an SPIB decoder predicts the metastable state of the system at a specific future time defined as the time-delay $\Delta t$. In this regard, SPIB can be thought of as a `fast mode filter' where the hyperparameter, $\Delta t$ can be used to tune the coarse-graining of the identified slow modes, as demonstrated in its original proof-of-principle publication.\cite{wang2021state}

Thus, for a given unbiased trajectory $\{\bm{X}^1,\cdots,\bm{X}^{M+s}\}$ and its corresponding state labels $\{\bm{y}^1,\cdots,\bm{y}^{M+s}\}$ with large enough $M$, we can employ the deep variational information bottleneck framework \cite{wang2021state,alemi2016deep} and construct an artificial neural network (ANN)  that is trained to maximize the following objective function:

\begin{equation} 
\begin{aligned} 
\label{eq:SPIB_obj}
\mathcal{L}=\frac{1}{M}\sum_{n=1}^M &\Bigl[\log q_{\theta}(\bm{y}^{n+s}|\bm{z}^{n})-\beta \log \frac{p_{\theta}(\bm{z}^{n}|\bm{X}^n)}{r_{\theta}(\bm{z}^{n})} \Bigr]
\end{aligned} 
\end{equation}
where $\bm{z}^{n}$ is sampled from $p_{\theta}(\bm{z}|\bm{X}^n)$ and the time interval between $\bm{X}^n$ and $\bm{X}^{n+s}$ is the time delay $\Delta t$, or how far into the future SPIB should predict. The first term $\log q_{\theta}(\bm{y}^{n+s}|\bm{z}^{(n)})$ measures the ability of our representation to predict the desired target, while the second term $\log \frac{p_{\theta}(\bm{z}^{n}|\bm{X}^n)}{r_{\theta}(\bm{z}^{n})}$ can be interpreted as the complexity penalty that acts as a regulariser. Such a trade-off between the prediction capacity and model complexity is then controlled by a hyper-parameter $\beta\in[0,\infty)$. All these three probability distributions $\left\{ p_{\theta}(\bm{z}|\bm{X}), q_{\theta}(\bm{y}|\bm{z}), r_{\theta}(\bm{z}) \right\}$ are implemented through deep neural networks with model parameters $\theta$. Further implementation details are provided in supplementary materials (SM).

However, for biased data generated from metadynamics, we need to reweight out the effect of the bias. Thus, along with the time series $\{\bm{X}^1,\cdots,\bm{X}^{M+s}\}$, we will also have the corresponding time-series for the bias $V$ applied to the system $\{V^1,\cdots,V^{M+s}\}$. We can then use the principle of importance sampling similar to Ref. \onlinecite{ribeiro2018reweighted} and write our objective function $\mathcal{L}'$ as follows:

\begin{equation} 
\begin{aligned} 
\label{eq:biased_SPIB_obj}
\mathcal{L}=\Bigl[\sum_{n=1}^M e^{\beta V^n}\Bigl]^{-1}\sum_{n=1}^M e^{\beta V^n}&\Bigl[\log q_{\theta}(\bm{y}^{n+s}|\bm{z}^{n})\\
&-\beta \log \frac{p_{\theta}(\bm{z}^{n}|\bm{X}^n)}{r_{\theta}(\bm{z}^{n})} \Bigr]
\end{aligned} 
\end{equation}
where $\beta$ is the inverse temperature. The above equation doesn't correct the kinetics for biased simulations, but as shown in Sec. \ref{sec:Results}, we found in practice it can still robustly  identify physically meaningful metastable states and high-quality RCs.



\subsection{System setup and OP description}
\label{sec:System_OP_details}

CHARMM36m\cite{huang2017charmm36m} all atom force field is used to parametrize the (Aib), and DMPC residues while CGenFF \cite{vanommeslaeghe2012automation} is used to parametrize BA. The (Aib)$_{9}$ system contains 4,749 atoms in total and is solvated using 1,539 TIP3P \cite{mackerell1998all,jorgensen1983comparison} water molecules. For the BA-DMPC system, the biological membrane is constructed as a phospholipid bilayer from 80 pure DMPC residues of which 40 DMPC forming upper and lower membrane leaflets respectively. The system contains a total of 222,785 atoms and is solvated using 71,102 TIP3P water molecules.

Since (Aib)$_{9}$ is a 9-residue peptide, a natural choice of OPs for this system, which is easily generalizable to other peptides, involves the nine $\phi$ and $\psi$ dihedral angles.\cite{biswas2018metadynamics} We consider sines and cosines of all the dihedrals to remove angular periodicity which amounts to a total of 36 OPs.

To construct an OP space for the BA-DMPC system, four different virtual position coordinates: the entire membrane center of mass (COM), membrane upper leaflet COM, membrane lower leaflet COM, BA benzene ring COM are first calculated. Based on these quantities, five key vectors from membrane COM to BA benzene ring COM ($\vec{d1}$), membrane COM to oxygen atom of BA ($-$OH) group ($\vec{d2}$), membrane COM to oxygen atom of BA (=O) group ($\vec{d3}$), BA benzene ring COM to oxygen atom of BA ($-$OH) group ($\vec{d4}$), lower leaflet COM to upper leaflet COM ($\vec{d5}$) are defined. $X, Y, Z$ components of $\vec{d1}, \vec{d2}, \vec{d3}$ constitute 9 OPs. Each of the 6 angles: $\theta_{X}, \theta_{Y}, \theta_{Z}, \omega_{X}, \omega_{Y},\omega_{Z}$ that $\vec{d4}$ and $\vec{d5}$ make with $X, Y, Z$ axes of the simulation box respectively are converted into sines and cosines to obtain 12 additional OPs. In total, a 21-dimensional input space for SPIB is constructed for BA-DMPC.

\subsection{SPIB augmented MD}
\label{sec:MD_details}

\begin{figure}
     \centering
         \includegraphics[width=\linewidth]{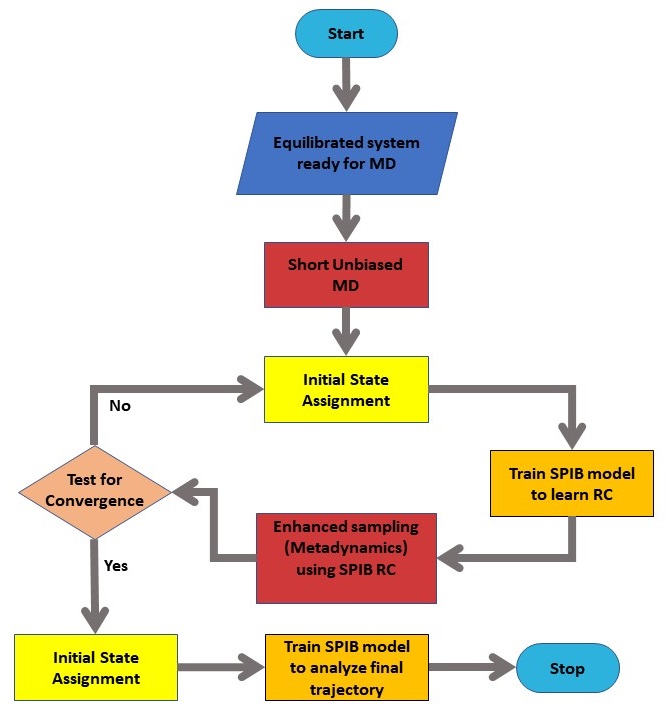}
         \caption{\textbf{Flowchart illustrates our protocol for SPIB-based enhanced sampling.} Starting with short unbiased MD trajectories, SPIB is employed to learn an optimal RC which to enhance the sampling of rare events. Well-tempered metadynamics \cite{valsson2016enhancing} was the enhanced sampling method employed in this work.}
         \label{fig:protocol}
     \hfill
\end{figure}

A complete workflow for applying SPIB to accelerate and interpret MD simulations is shown in Fig. \ref{fig:protocol}. Its key aspects are exemplified below in the context of the two systems studied here: (1) conformational transitions in (Aib)$_{9}$ peptide and (2) BA permeation through phospholipid bilayer.

The starting point of this protocol involves performing a relatively short unbiased MD simulation that provides time-series of the features or OPs. Training an ANN model using SPIB involves feeding these OPs and the initial state assignments with minimal use of human intuition to construct a low dimensional system RC.

The initial assignment of states can be carried out in at least two different robust manners with their respective strengths and weaknesses (see Sec. \ref{sec:Results} and SM). If one has \textit{a priori} structural information about the system, then this can be directly used in what we call a ``structural" scheme. However, if no such information is known at all, one can simply partition the time-series into discrete labels \cite{hyvarinen2016unsupervised} which then serve as a ``temporal" scheme for initial state assignment.

The RC is then utilized to perform well-tempered metadynamics and achieve enhanced sampling. Ideally, the length of the unbiased simulation used to learn system RC, should be long enough so that the rare event of interest has been seen at least once. However, for truly rare event systems such as ligand dissociation \cite{shekhar2021protein,PANT2020} this might be impossible, and in this case the protocol might need more rounds to converge to an optimized RC as shown in Fig. \ref{fig:protocol}. The metadynamics trajectory, appropriately reweighted, can be used to learn an improved RC and corresponding metastable states. In principle, this trajectory RC can also be used to perform further metadynamics to obtain even better sampling \cite{ribeiro2018reweighted,wang2019past,wang2021interrogating} as we do for the (Aib)$_{9}$ system (Sec. \ref{sec:Results}). Final metadynamics trajectories for both systems, with reweighting factors accounting for the bias \cite{tiwary2015time} are again analyzed through SPIB to identify the final RC and relevant metastable states which we report in Sec. \ref{sec:Results}. All MD simulations are performed by employing Nose-Hoover thermostats and Parrinello-Rahman barostats \cite{parrinello1980crystal, hoover1985canonical} using GROMACS 2020.2 \cite{van2005gromacs}, patched with PLUMED\cite{tribello2014plumed} 2.6.2. Further simulation details can be found in the SM.

\section{Results}
\label{sec:Results}

\subsection{Chiral transitions in (Aib)$_{9}$}
\label{sec:aib_result}

\begin{figure*}[t!]
    \centering
    \includegraphics[width=\linewidth]{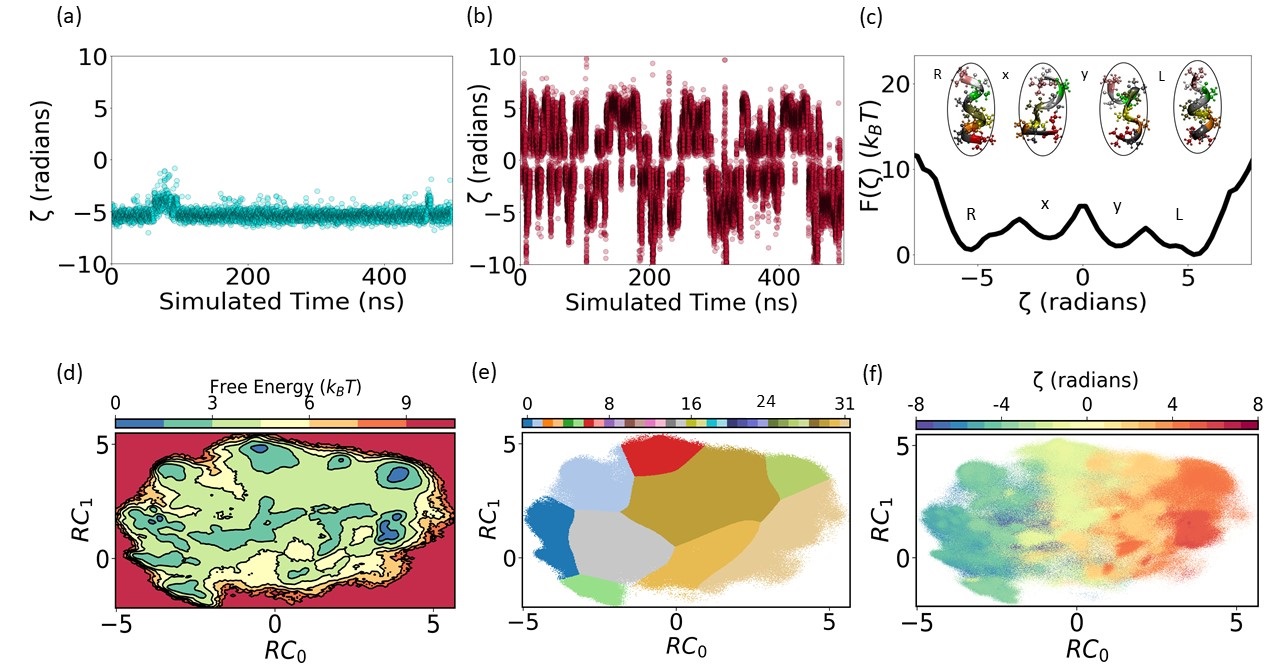}

    \caption{\textbf{Chiral transitions in (Aib)$_{9}$.} $\zeta$ time series for (a) unbiased MD, and (b) SPIB augmented metadynamics. Multiple back and forth transitions were captured in SPIB metadynamics compared to unbiased MD. (c) Free energy along OP $\zeta$ highlighting R,rlrrl, llrlr, L metastable states showing similar free energy for L and R, (d) free energy along 2-d RC for time delay, $\Delta t = 0.3 ns$, (e) converged state labels, and (f) previously used\cite{biswas2018metadynamics} expertise-based RC $\zeta$ projected on the 2-d RC generated by SPIB.}
    \label{fig:time_series} 
\end{figure*}

We stay consistent with Ref. \onlinecite{biswas2018metadynamics} and define the fully left (L) handed and fully right (R) handed configurations of (Aib)$_{9}$ only based on the inner 5 amino acid residues. The system transitions from L to R and vice-versa as the chirality of each of the inner 5 residues flip from one to the other. In subsequent discussions we implemented ``structural" initial state assignment scheme for SPIB by discretizing along the dihedral angle $\phi$ of the inner 5 residues, leading to $2^5=32$ initial state labels. We also explore an alternate approach for (Aib)$_9$ which we call `temporal' initial state assignment scheme and relevant results for this approach are provided in SM.

At first, we conducted a short $500 ns$ unbiased MD for (Aib)$_9$ in NPT ensemble at 400K and 1 atm. However, this is not long enough to observe R-L transitions and the system remained trapped in the R helical conformation during the entire trajectory as demonstrated in Fig. \ref{fig:time_series}(a) by the hybrid OP $\zeta \equiv \sum_{n=3}^{7}\phi$, which we refer to as expertise-based RC.\cite{biswas2018metadynamics} As $\phi\approx-1$ for a right-handed residue and $\phi\approx1$ for a left-handed residue, $\zeta\approx-5$ and $\zeta\approx5$ correspond to the fully right (R) handed and fully left (L) handed states respectively.

From such an under-sampled unbiased trajectory, a 2-dimensional RC is learned by SPIB and used as the biasing variable to perform $200 ns$ metadynamics simulation. Even in this first round, we observe many new metastable states and a full back and forth transitions between the two most stable states (L and R) (see SM). This biased trajectory is reweighted and used to run a second round of SPIB to determine a more informative RC. Subsequently, this improved RC is used as the biasing variable of metadynamics to perform $700 ns$ MD simulation. This trajectory contains multiple back and forth transitions between the two most stable states (L and R) as shown in Fig. \ref{fig:time_series}(b). Therefore, we finally achieve an acceleration of 40 times by considering number of back and forth per unit time after 2 rounds of SPIB-metadynamics. Fig. \ref{fig:time_series}(b) highlights the first $500ns$ metadynamics and the complete $700ns$ simulation result is provided in SM.

Fig. 3(c) demonstrates the free energy along $\zeta$ highlighting two intermediate metastable state regions x, and y between R and L taking the reweighted metadynamics into account.\cite{tiwary2015time} The rlrrl, and llrllr (Aib)$_9$ conformations are highlighted that belong to these intermediate metstable states respectively.This figure clearly shows the limitations of such expertise-based RC, as there is a huge degeneracy of conformations along $\zeta$ and it fails to provide a clear picture of (Aib)$_9$ chiral transitions.

To interpret this $700 ns$ MD trajectory, a third round of final SPIB models were trained to learn system RC and gain biophysical insights. While training SPIB models, the SPIB hyperparameter, $\Delta t$ was used to neglect the fast modes. A range of different $\Delta t$ were chosen while training different models and the number of converged states detected by SPIB decreased with $\Delta t$ (see SM). Fig. \ref{fig:time_series}(d) shows free energy along the 2-d RC space for a particular $\Delta t = 0.3ns$ that detects a reasonable number of metastable states. The trained SPIB model detected 9 converged state labels for $\Delta t = 0.3 ns$. In Fig. \ref{fig:time_series}(e), fully right (R) handed and fully left (L) handed states are classified to SPIB states 0 and 31 respectively, while the intermediate states between R and L are represented by the other 7 converged states. The remaining 23 SPIB states were found to have no significant population after training convergence. Projection of $\zeta$ on this 2-d RC space clearly demonstrates the strong correspondence between our RC learned by SPIB and the expertise-based RC ($\zeta$) as shown in Fig. \ref{fig:time_series}(f). Here, $RC_0$ is similar to $\zeta$ as it differentiates between L and R metastable states while $RC_1$ distinguishes between the metastable states that are degenerate in $\zeta$.

\subsection{BA permeation through phospholipid bilayer}
\label{sec:BA_result}

For this system we are guided by two simple pieces of intuition. Firstly, a successful permeation event can occur when BA explores the direction of membrane surface normal. Secondly, the polar region of BA possibly plays a role in permeation process by interacting with the membrane. Based on this intuition, initial states for BA-DMPC were assigned by using ${d1_Z}$, and $\theta_Z$. To demonstrate that SPIB is capable of utilizing independent and discontinuous MD trajectories, we learnt a 1-d RC by combining two $25 ns$ unbiased trajectories launched by initially placing BA on the two opposite sides of the membrane along Z direction. In both simulations, BA gets trapped near the membrane surface regions after entering the membrane (see SM).

\begin{figure*}[t!]
    \centering
    \includegraphics[width=\linewidth]{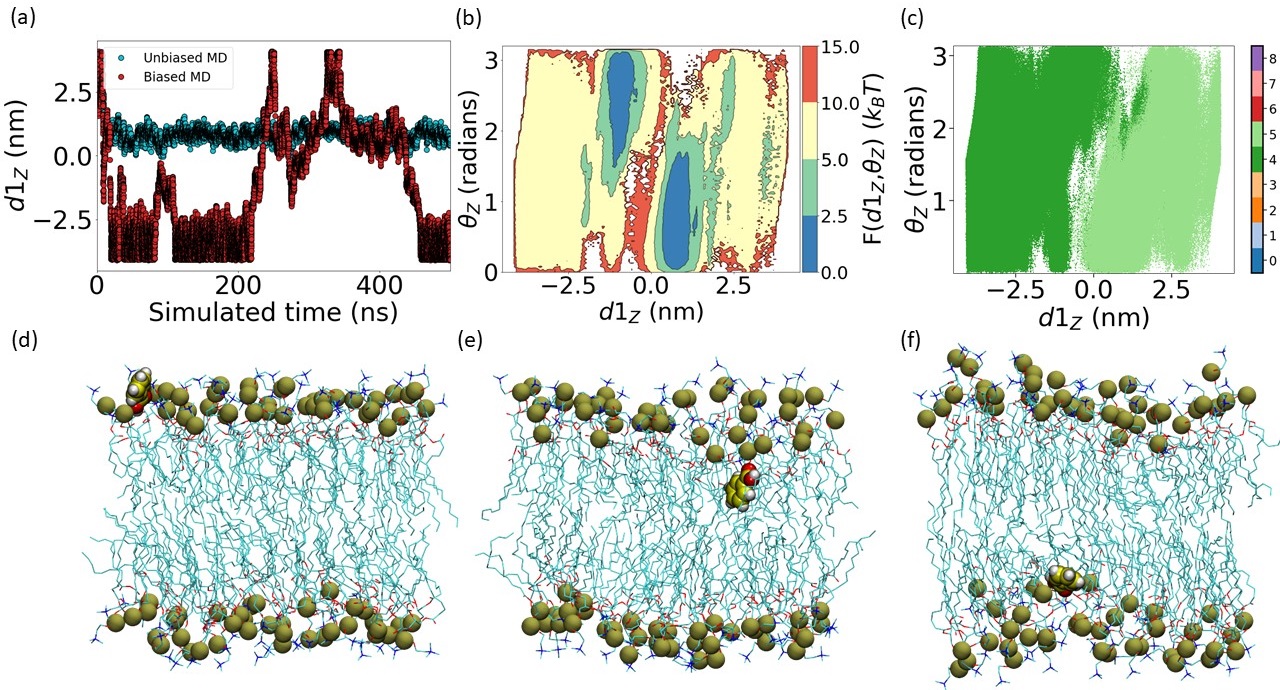}

    \caption{Time series data for (a) $500ns$ unbiased MD started from a configuration with BA placed outside the membrane , and SPIB based metadynamics, (b) free energy along ($d1_Z, \theta_Z$) for biased simulation. (c) two converged SPIB states depicting key permeation barrier. Permeation mechanisms are highlighted in (d, e, f) where in respective cases, BA has found an entry point into the membrane, is stuck at the surface along +Z direction, and is stuck at the surface along -Z direction from membrane COM.}
    \label{fig:BA-DMPC} 
\end{figure*}

$500 ns$ metadynamics based on this 1-d RC achieves 3 complete permeation events compared to none in unbiased MD as shown in Fig. \ref{fig:BA-DMPC} (a). Free energy along ($d1_Z,\theta_Z$) highlights the tendency of BA to stay trapped near membrane surface region upon entry (Fig. \ref{fig:BA-DMPC}(b)). The position of the key barrier acting against this permeation process was identified by the SPIB learnt 1-d RC as shown in Fig. (\ref{fig:BA-DMPC}(c)). A relatively high time delay, $\Delta t=20 ps$ was chosen to recognize two metastable states and thus identifying the key barrier.

Additionally, SPIB augmented metadynamics enabled the study of interesting entry/exit mechanisms of BA shown in Fig. \ref{fig:BA-DMPC}(d,e,f). Prior to entering the membrane, the polar region of BA begins interacting with membrane surface and acts as an anchor. This enables the BA benzene ring to flip inside the membrane and enter. Once inside, the BA polar region keeps interacting with the headgroup region of the membrane and the hydrophobic BA benzene ring positions itself close to membrane center, where it interacts with the lipid tails. Finally, after overcoming the key barrier, which corresponds to its flipping within the membrane, BA reaches the other leaflet of the membrane and BA polar region starts interacting with membrane  headgroup. Thus, ligand permeation here is a composite of entropic and enthalpic barriers, corresponding respectively to diffusion to the membrane and subsequent escape beyond the membrane surface. Notice that metadynamics helped here with speeding up the movement across the energetic barrier significantly (Fig. \ref{fig:BA-DMPC}(a)). The entropic process could also be improved further by using SPIB in a method such as weighted ensemble, milestoning or forward flux sampling. \cite{zuckerman2017weighted, faradjian2004computing,defever2019contour}

It can be argued that a traditional metadynamics with $d1_Z$ as the biasing variable can also generate membrane permeation events. However, SPIB learnt 1-d RC learns the underlying physics and takes contributions from all the 21 OPs into account including $d1_Z$. For example, Fig. \ref{fig:BA-DMPC}(c) exemplifies that SPIB considered $\theta_Z$ contributions in addition to $d1_Z$ for identifying the key barrier. In this way, the resultant trajectory will reflect biophysically relevant events and avoid brute-force entry that would possibly be observed when using only $d1_Z$ as the biasing variable.\cite{lee2016simulation} This demonstrates a key strength of SPIB augmented metadynamics.

\section{Conclusion}

In this work, we have applied the recently developed SPIB framework\cite{wang2021state} of the RAVE class of methods\cite{ribeiro2018reweighted,wang2019past} to accelerate and understand two prototypical biophysical problems plagued with rare events that are inaccessible in microsecond-long unbiased MD. The two problems considered here are protein conformational dynamics in a 9-residue peptide (Aib)$_9$\cite{sittel2017principal, biswas2018metadynamics} and permeation of the small molecule benzoic acid through a phospholipid bilayer. For both systems, the SPIB reaction coordinate based metadynamics was able to successfully and significantly accelerate the simulations. In addition, the SPIB augmented metadynamics helped us gain physical insights into the respective problems. For (Aib)$_{9}$, SPIB identifies the complex energetic landscape behind chiral transition process and for BA-DMPC, discovers the position of the key permeation barrier. In future, we aim to augment SPIB with other enhanced sampling schemes to effectively tackle bio-physical problems across spatio-temporal scales. \newline

\textbf{Supplementary material\newline }
See supplementary material for system information, neural network architecture and other details. \newline

\textbf{Acknowledgements\newline }
Research reported in this publication was supported by the National Institute Of
General Medical Sciences of the National Institutes of
Health under Award Number R35GM142719 (P.T.)
The content is solely the responsibility of the authors
and does not necessarily represent the official views
of the National Institutes of Health. The authors thank Deepthought2, MARCC, and
XSEDE (projects CHE180007P and CHE180027P) for
providing computational resources used in this work. The authors thank Dr. Eric Beyerle, Yihang Wang and Zachary Smith for helpful discussions and valuable insights. \newline

\textbf{Data availability statement\newline }
The data and codes that support the findings of this study will be made available through GitHub and PLUMED-NEST. \newline

\textbf{Conflict of Interest\newline}
The authors declare the following competing financial
interest(s): P.T. is a consultant to Schrodinger, Inc. and S.P. is currently an employee of Loxo Oncology {@} Lilly and is a shareholder of stock in Eli Lilly and Co. \newline \newline

\textbf{References}

	\end{document}